# Novae ejecta as colliding shells

**Robert Williams[1] and Elena Mason[2]**

[1]Space Telescope Science Institute, 3700 San Martin Drive, Baltimore, MD 21218
[2]European Southern Observatory, Alonso Cordova 3107, Vitacura, Santiago, Chile

E-mail: wms@stsci.edu

**Abstract:** Following on our initial absorption-line analysis of fifteen novae spectra we present additional evidence for the existence of two distinct components of novae ejecta having different origins. As argued in Paper I one component is the rapidly expanding gas ejected from the outer layers of the white dwarf by the outburst. The second component is pre-existing outer, more slowly expanding circumbinary gas that represents ejecta from the secondary star or accretion disk. We present measurements of the emission-line widths that show them to be significantly narrower than the broad P Cygni profiles that immediately precede them. The emission profiles of novae in the nebular phase are distinctly rectangular, i.e., strongly suggestive of emission from a relatively thin, roughly spherical shell. We thus interpret novae spectral evolution in terms of the collision between the two components of ejecta, which converts the early absorption spectrum to an emission-line spectrum within weeks of the outburst. The narrow emission widths require the outer circumbinary gas to be much more massive than the white dwarf ejecta, thereby slowing the latter's expansion upon collision. The presence of a large reservoir of circumbinary gas at the time of outburst is suggestive that novae outbursts may sometime be triggered by collapse of gas onto the white dwarf, as occurs for dwarf novae, rather than steady mass transfer through the inner Lagrangian point.

*Subject headings*: Novae, cataclysmic variables
*Short Title:* Novae Colliding Shells

## 1. Introduction

The geometry and kinematics of novae ejecta are central to the interpretation of their spectra following the outburst. Most models indicate that nuclear reactions continue at a much reduced rate following the outburst spike, and that these drive a wind (Kato & Hachisu 2009) that may possibly be preceded by the ejection of a portion of the outer layers of the white dwarf (Starrfield, Truran, & Sparks 2000; Yaron et al. 2005). The difficulty in treating the hydrodynamics of the brief period (seconds) when the temperature spikes and the partially degenerate gas becomes non-degenerate has led to considerable uncertainty in the extent of episodic mass ejection associated with the thermonuclear runaway (TNR). Some novae are observed to have a luminosity that exceeds the Eddington limit for weeks after the outburst, driven by continued nuclear burning (Schwarz et al. 2001; Kato & Hachisu 2007), and detailed radiative transfer models of dense photospheric winds by Hauschildt et al. (1997) and Short et al. (2001) using the PHOENIX code have found good fits to both the UV and optical postoutburst spectra for a range of abundances and temperature-density laws.

A specific view of mass ejection from the outburst was proposed by Williams et al. (2008, hereafter called Paper I) based on time sequences of high resolution visible spectra of novae in the months following outburst. They observed multiple absorption systems having different radial velocities in the spectra of a large majority of novae and interpreted the observations in terms of discrete clouds of gas that are progressively accelerated outward. The innermost ejecta, having the higher velocities, correspond to the surface layers of the white dwarf (WD) that have undergone nuclear reactions. The outermost ejecta are circumbinary gas having lower velocities, and their presence at maximum light surrounding the primary WD ejecta strongly suggests that the outer circumbinary transient heavy element absorbing (THEA) gas has existed before the TNR.

Based on the abundances and velocities of the absorption systems Paper I proposed that an outer circumbinary gas reservoir having mass of order $10^{-5}$ $M_\odot$ is ejected episodically by the secondary star prior to the outburst. A fraction of the ejected material escapes the system and a fraction remains bound to the system, accreting onto the WD and triggering the nova outburst. K. Nomoto and M. Shara (2008, private communications) have called attention to the fact that the energy required to produce the circumbinary reservoir is of order $10^{44}$ ergs, i.e., roughly the total energy released in a classical nova outburst. Thus, an individual energetic event the magnitude of a typical outburst is required to create the putative pre-outburst circumbinary gas, and would be expected to rival the main outburst in luminosity. Yet, no separate precursor event of this magnitude is observed for novae.

We suggest here a slightly modified scenario based on an analysis of the emission line profiles from our previous survey spectra and on the hydrodynamical calculations of Sytov et al. (2007, 2009b) which show that a circumbinary reservoir is created in cataclysmic variables by steady mass loss from the outer accretion disk through the outer L3 Lagrangian point. This hypothesis is testable through the detection of the circumbinary reservoir via Na I and Ca II absorption in quiescent nova systems. We show that postoutburst novae absorption and emission line widths and profiles are consistent with a picture in which the fast moving WD outburst ejecta collide with a significantly more massive outer reservoir of gas, and that radiation from the outburst can account for the observed acceleration of both components of expanding gas.

## 2. Novae ejecta

*2.1. Circumbinary gas characteristics*

In his original spectral classification scheme for novae McLaughlin (1943, 1947) assumed that the multiple absorption systems having different properties and radial velocities originated from distinct shells of gas. Independently, H. N. Russell (1936) and S. Rosseland (1946) suggested that discrete shells might be created by the shock wave from the outburst, so the concept of separate shells of gas ejected by novae is not new. Their existence has been reinforced by the analyses of IUE high resolution UV spectroscopic data of the nova V1974 Cygni/1992 (Cassatella et al. 2004) and the high resolution optical spectra from Paper I. Arguments were given in Paper I that the origin of the circumbinary gas is the secondary star, possibly via the outer accretion disk, based upon the presence of Fe-peak elements and the lower expansion velocities of the systems, which are similar to the escape velocities of the secondary stars. The energetics of creating a pre-maximum component of gas favors its origin on either the secondary star or accretion disk since the potential wells of the disk and the secondary are much shallower than that of the WD.

An approximate spectral analysis of one outer gas shell was presented in Paper I under simplified assumptions for the nova LMC 2005, which showed the most well defined spectrum of such systems. There are several other novae in our high resolution optical survey for which some information might be obtained from the observed THEA systems, and we have performed a similar analysis of these absorption systems under the same assumptions since the conditions under which the lines are formed are not known. With the exception of the novae LMC 2005 and V378 Ser/05 the THEA lines are so broad in these other systems that line blending is a serious problem, as is clear from figure 2 of Paper I. Nevertheless, some information is available from individual multiplets that appear to be reasonably well defined. The relative equivalent widths of the lines within a multiplet reveal potential saturation of the lines, e.g., whether column densities derived from such lines represent lower limits.

We have measured the equivalent width of lines in five THEA systems that by their widths and profiles should be unblended. For lack of more specific information we assume the lines to be formed under ISM conditions of low density in a dilute radiation field, and we use the data to determine the excitation temperature and ion column densities (Jenkins et al. 2005). We present the results of the measurements and the derived parameters for the five novae in table 1 together with the f-values of the lines. The analysis assumes the lines to be optically thin, so when equivalent widths of other lines in the same multiplet do not have the same ratios as the quantity ($gf\lambda^2$) they are either blended or saturated lines. The parameters derived from such lines are not reliable and are marked with a colon, and actual column densities will be greater than those calculated on the basis of assumed optically thin lines. In particular, the excitation temperatures derived from the Sc II lines are anomalously high, due either to poor f-values or blends.

Table 1. THEA system equivalent widths and column densities

| Line | log (gf) | σ (cm⁻¹) | V378 Ser/05 5 April 2005 | | | V5116 Sgr/05#2 7 July 2005 | | | V2573 Oph/03 15 August 2003 | | |
|---|---|---|---|---|---|---|---|---|---|---|---|
| | | | EW (Å) | FWHM (km/s) | log N (cm⁻²) | EW (Å) | FWHM (km/s) | log N (cm⁻²) | EW (Å) | FWHM (km/s) | log N (cm⁻²) |
| Ti II 4012.38 | -1.61 | 4629 | 0.081 | 28 | | | | | | | |
| Sr II 4077.71 | 0.17 | 0 | | | | 0.17 | 57 | 12.4 | | | |
| Ti II 4163.65 | -0.40 | 20892 | 0.19 | 44 | | 0.35 | 95 | | 0.69 | 90 | |
| Sr II 4215.52 | -0.14 | 0 | | | | 0.09 | 57 | | | | |
| Sc II 4246.82 | 0.32 | 2541 | 0.18 | 35 | 13.4 | 0.25 | 74 | 13.6 | 0.33 | 87 | 13.7 |
| Ti II 4294.10 | -1.11 | 8744 | 0.12 | 26 | 15.4 | 0.33 | 89 | 15.9 | 0.46 | 103 | 16.0 |
| Sc II 4325.00 | -0.44 | 4803 | | | | | | | | | |
| Ti II 4411.07 | -1.06 | 24961 | | | | | | | | | |
| Sc II 4670.40 | -0.37 | 10945 | | | | | | | | | |
| Sc II 5526.79 | 0.13 | 14261 | 0.11 | 37 | | 0.10 | 76 | | 0.36 | 94 | |
| $T_{exc}$(K) | | | 11,675 (Ti II) 29,010 (Sc II) | | | 11,560 (Ti II) 16,800 (Sc II) | | | 15,000 (Ti II) >10⁶ (Sc II) | | |

| Line | log (gf) | σ (cm⁻¹) | V1186 Sco/04#1 9 August 2004 | | | V5117 Sgr/06 21 February 2006 | | |
|---|---|---|---|---|---|---|---|---|
| | | | EW (Å) | FWHM (km/s) | log N (cm⁻²) | EW (Å) | FWHM (km/s) | log N (cm⁻²) |
| Ti II 4012.38 | -1.61 | 4629 | 0.12 | 28 | | 0.18 | 46 | 15.9 |
| Sr II 4077.71 | 0.17 | 0 | 0.093 | 31 | 12.2 | 0.29 | 66 | 12.7 |
| Ti II 4163.65 | -0.40 | 20892 | 0.15 | 36 | | | | |
| Sr II 4215.52 | -0.14 | 0 | 0.11 | 25 | | 0.18 | 52 | |
| Sc II 4246.82 | 0.32 | 2541 | | | | 0.20 | 58 | 13.5 |
| Ti II 4294.10 | -1.11 | 8744 | 0.12 | 36 | 15.4 | | | |
| Sc II 4325.00 | -0.44 | 4803 | 0.094 | 37 | 14.0 | | | |
| Ti II 4411.07 | -1.06 | 24961 | 0.056 | 31 | | 0.053 | 35 | |
| Sc II 4670.40 | -0.37 | 10945 | | | | 0.16 | 34 | |
| Sc II 5526.79 | 0.13 | 14261 | 0.085 | 33 | | 0.23 | 48 | |
| $T_{exc}$(K) | | | 17,000 (Ti II) 7,200 (Sc II) | | | 10,940 (Ti II) >10⁶ (Sc II) | | |

The results in table 1 show that the parameters derived for Nova LMC 2005 in Paper I are typical of novae, in general. The circumstellar gas has temperatures of ~$10^4$ K, characteristic of ionized gas. And the column density of the absorbing systems, assuming roughly solar abundances, is of the order of ~$10^{23}$ cm$^{-2}$ for the THEA systems we are able to analyze. We emphasize that these results are only approximate because of our lack of knowledge of the conditions under which the lines are formed.

The large column densities of the THEA systems raise the possibility that dust may form in the outer circumbinary gas. The existence of dust in this gas cannot be ruled out even though the ionization and temperatures of $T \sim 10^4$ K for the gas near the time of the outburst are not favorable for dust formation. Dust is known to exist within the ionized gas of HII regions and planetary nebulae, which have similar conditions, so it could be present in the THEA gas. Dust is more likely to exist in the CNO enhanced, rapidly expanding adiabatically cooled WD ejecta (see Shore & Gehrz 2004). Its presence in either type of absorbing system would contribute significantly to the radiative acceleration of the systems because of large dust absorption cross sections for radiation.

Since the THEA systems disappear as forbidden lines emerge, the absence of a UV continuum in the earlier photospheric phase prevents narrow THEA absorption from being detected in the UV for most novae. THEA absorption is thus confined primarily to optical wavelengths, and relatively little information about it is available in the UV. There are a few exceptions, however. Cassatella et al. (2004) studied IUE archival spectra of one nova, V1974 Cygni/1992, which showed both narrow and broad ultraviolet absorption features for a period of weeks following the outburst. Narrow UV THEA absorption lines were observed from low ionization species Fe II, Si II, and Al II, as were strong broad, higher expansion velocity P Cygni absorption features from both high and low ionization resonance lines such as C IV λ1549, N V λ1240, Al III λ1857, C II λ1335, and O I λ1302. These UV data confirm that the prominent optical Na I D and Ca II H&K P Cygni features observed in novae are the low ionization component of the expanding primary ejecta, which have a wide range in ionization.

*2.2. Radiative acceleration of ejecta*

Outburst calculations by different groups show that the ejection of $\sim 10^{-5}$ $M_\odot$ of the surface layers of the WD at velocities exceeding $10^3$ km/s is a general outcome of the outburst (Glasner, Livne, & Truran 2007; Kato & Hachisu 2009). The spectral signatures of the ejecta are the broad P Cygni absorption features in the optical Balmer, Fe II, and Na I D lines and in the CNO UV resonance lines, all indicating an expansion velocity of order $2\text{-}4 \times 10^3$ km/s and internal velocity dispersion >1,000 km/s. A second component of absorbing gas is the circumstellar reservoir, having lower expansion velocities (400-1,000 km/s) and smaller internal velocity dispersion (35-300 km/s). Its spectral signature is the concentration of heavy element absorption lines between 3800-5600 Å. The multiple systems of expanding gas caused by the outburst present two energy considerations: (1) constraints on the ejection of circumbinary gas from the binary that are imposed by the total outburst energy, and (2) postoutburst energy generation requirements that account for the observed progressive acceleration of the absorbing gas.

The majority of the thermonuclear energy produced in the nova outburst is in the form of MeV gamma rays emitted by proton-capture reactions and in $\beta^+$ particles given off by the proton-rich nuclei as they radioactively stabilize. The positrons immediately annihilate with electrons and produce 511 keV gamma rays. The gamma rays Compton scatter, progressively down-scattering and losing energy to the heating of free electrons. A column density of order $10^{25}$ cm$^{-2}$ is required to scatter most of the γ-rays so they are converted into X-rays by the end of the process (Senziani et al. 2008; Orio 1999). Models have not yet resolved the question whether the TNR takes place sufficiently deep in the outer layers of the WD that the majority of gamma rays are converted to X-rays before the radiation escapes the region where it is created (Gomez-Gomar et al. 1998). As the primary source of energy released in the outburst the radiation is most likely to be responsible for the ejection and acceleration of matter from the binary. A desirable feature of radiative acceleration of the ejecta is that it can accelerate the gas coherently and compressively (Williams 1972), thus accounting for the small internal velocity dispersions observed in some THEA systems.

The energy produced in the thermonuclear runaway, $E_{nuc} \sim 10^{45}$ ergs (Yaron et al. 2005), must be sufficient to lift circumbinary gas from a distance r out of the binary potential well and give it an ejection velocity of order $v_f$ ~700 km/s. Assuming the gas to absorb a fraction of the radiation, energy conservation yields

$$\alpha\, E_{nuc} = GMm_{cir}/r + \tfrac{1}{2}\, m_{cir} v_f^2 , \qquad (1)$$

where M is the combined mass of the two stars. The numerical coefficient α, which takes on values of ~0.1-10, accounts for the fact that (1) there is likely to be incomplete absorption of the radiation by the gas, and (2) radiation, which communicates the outburst energy to the gas, can scatter multiple times before escaping, imparting momentum to the gas on each scattering. We will assume a circumbinary mass of $m_{cir}=10^{-5}$ $M_\odot$, which is the mass determination from Paper I for the system analyzed there, and the conservative condition that α~1. For a combined binary system mass of one solar mass equation (1) then requires the circumbinary gas to be located at the time of outburst at $r > 3\times 10^9$ cm from the stars. This criterion is met if it originates near the secondary star or the outer accretion disk. However, if the circumbinary gas originates deeper in the potential well near the WD this condition is not met. Mass loss from the secondary is energetically more favorable and satisfies the energy constraint as long as the gas absorbs a significant fraction of the radiation. Complete absorption of the gamma rays created in the outburst requires a column density that is larger than the $10^{23}$ cm$^{-2}$ derived in Paper I for Nova LMC 2005 and in §2.1 above for other THEA systems. On the other hand, if the radiation escapes the expanding WD photosphere predominantly as X-rays the circumbinary column densities are sufficient to absorb most of the radiation (Livio et al. 1992; Fruchter, Krolik, & Rhoads 2001).

A second energy constraint concerns the observed acceleration of THEA systems in the weeks after maximum light before they disappear, requiring a continuing source of energy to drive the acceleration. Examples of this acceleration have been shown in figures 1 and 3 of Paper I, with the latter figure giving quantitative velocity information on the acceleration. We display several additional examples in figure 1 for the Na I D region of the spectrum. The simultaneous acceleration of both the primary ejecta (broad Na I D P Cygni absorption) and the THEA systems (narrower, resolved Na I D doublets) is evident in these figures. Without a sustaining source of energy maintaining external pressure on the gas the velocities of both ejecta systems would remain constant and perhaps even decrease. An adequate source of energy is required to explain the acceleration.

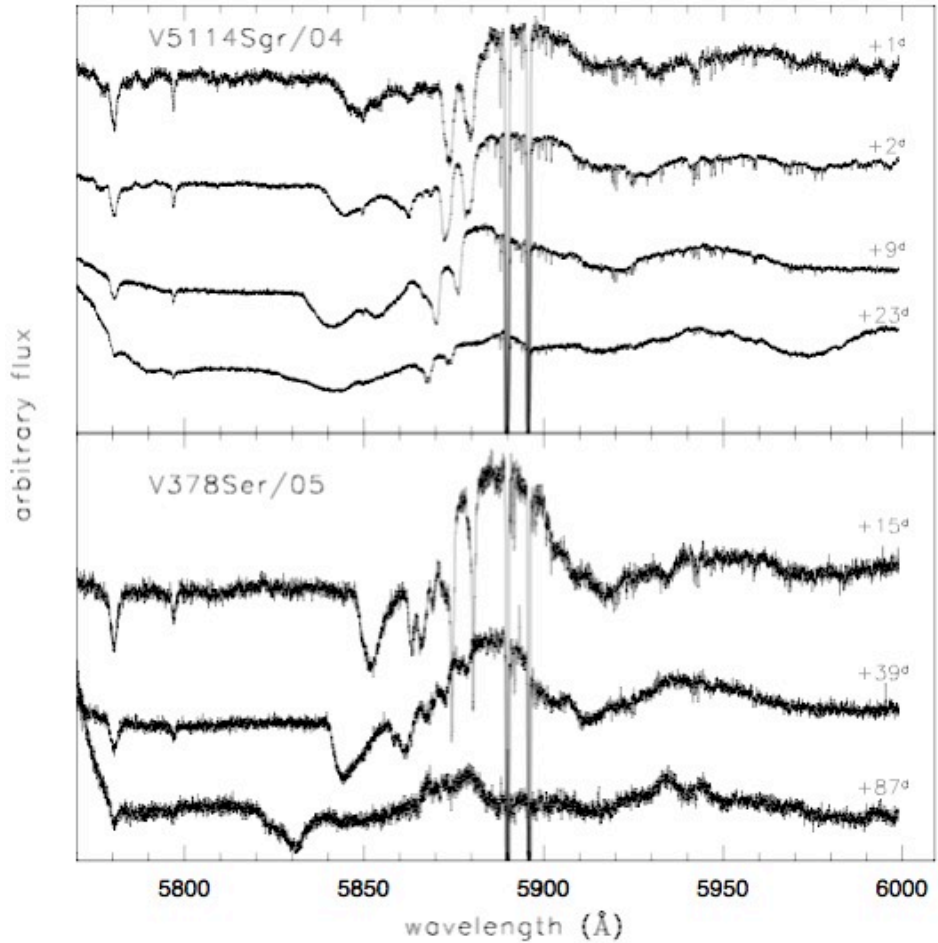

**Figure 1.** Time sequence of spectra of the novae (a) V5114 Sgr/04, and (b) V378 Ser/05, showing the acceleration of both the broad, unresolved primary ejecta P Cygni Na I D absorption (blueward of 5855 Å) and the narrower, resolved Na I D THEA absorptions. The absorption features at 5780 & 5797 Å are diffuse interstellar bands.

The radiation pressure, i.e., force per unit area, f, acting on matter that absorbs radiation having flux $F=L_{rad}/(4\pi r^2)$ is $f = F/c$ (Mihalas 1978). The momentum imparted to a column of gas absorbing the radiation produces an acceleration $\Delta v/\Delta t$ set by the condition

$$L_{rad}/(4\pi c r^2) = m_H\, N\, \Delta v/\Delta t, \qquad (2)$$

where N is the column density of the absorbing gas located a distance r from the radiating source.

It has been determined, especially for LMC novae whose distances are known, that some novae maintain a super-Eddington luminosity for periods of weeks following the outburst (Schwarz et al. 2001; Kato & Hachisu 2007). The THEA absorption systems detected in our survey were observed in the weeks following maximum light to undergo accelerations ranging in values from 0 km/s per day, i.e., no acceleration (LMC 2005, V382 Nor, V476 Sct), to a maximum of 30 km/s per day (V5114 Sgr), with a median value of ~5 km/s per day. The large majority of THEA systems do not experience any acceleration in excess of 5 km/s per day. The column densities of the THEA systems we have analyzed are all in the vicinity of $N \approx 10^{23}$ cm$^{-2}$. Assuming the novae to maintain a luminosity near the Eddington limit for a one solar mass star, $L_{edd} = 2 \times 10^{38}$ erg/s, equation (2) requires THEA system acceleration to occur within a distance of $r < 2.3 \times 10^{13}$ cm ~2 AU. This limit is greater than the distance THEA gas travels expanding ~700 km/s in the time between the outburst and our early observations. Thus, radiation pressure is a viable mechanism to explain the observed acceleration of ejecta in novae.

**3. Interaction of ejecta**

The different characteristics and evolution of the various absorption features of spectral lines is the strongest argument for the existence of separate components of ejecta. The primary WD ejecta give rise to what McLaughlin termed the 'diffuse enhanced' spectrum, having expansion velocities exceeding 2,000 km/s. They are generally enriched in CNO (Gehrz et al. 1998) and have a wide range of ionization and internal velocities as evidenced by the broad P Cygni profiles of resonance lines from Na I to N V.

The second component of ejecta is the circumbinary gas, which exists outside the primary nova ejecta and does not show the wide dispersion in properties evident in the primary ejecta. This reservoir of gas gives rise to the transient heavy element absorption THEA systems that McLaughlin classified as the 'principal' spectrum. THEA systems have outward velocities generally lower than 1,000 km/s, low ionization, and internal velocity dispersions ranging from 35-300 km/s.

The circumbinary gas must be very clumpy and inhomogeneous as evidenced by the fact that the THEA absorption lines indicate saturation, as discussed in §2.1, and yet the residual intensities of the saturated line centers are always greater than 0.5 times the intensity of the neighboring continuum, never approaching zero intensity (see figure 3 of Paper I). This behavior requires THEA gas to have a covering factor of order 0.5. Since THEA absorption is observed in 80% of novae the gas must exist along most lines of sight, i.e., have a roughly spherical distribution, and be very 'patchy', with individual clumps being smaller than the expanding central continuum source. This interpretation is supported by the observation in novae of multiple narrow THEA systems having different radial velocities, e.g., some novae have as many as 4-5 distinct Na I D absorption doublets, each having an equivalent width ratio indicating line saturation but with central line intensities generally exceeding 0.5 that of the continuum intensity (figure 1 of Paper I).

An important consequence of the two components of ejecta having very different velocities is that the inner, higher velocity primary WD ejecta must eventually collide with the outer, more slowly expanding circumstellar gas, and the interaction of these two components will dictate the subsequent evolution of the spectrum. This situation is analogous to the expanding interstellar bubbles driven by the winds of hot stars (Castor, McCray, & Weaver 1975). Depending upon the relative velocities of the two components and the location of the circumbinary gas at the time of the outburst the timescale for the collision to happen should be of order 2-6 weeks. This is the same timeframe over which we observe most novae to change from the initial absorption spectrum to a prominent forbidden emission-line spectrum.

The physics of the interaction of a hot star wind impacting interstellar gas has been treated by Castor, McCray, & Weaver (1975) and Weaver et al. (1977). In the context of novae the role of the wind is assumed by the primary WD ejecta, and the circumbinary gas is the proxy for the ISM. The primary ejecta form an expanding photosphere that cools rapidly (Arnett 1979) and radiates the observed broad P Cygni spectrum, with most of the emission component of the P Cygni profiles being due to scattering in the optically thick ejecta. The outer circumbinary THEA gas is warm and has a low level of ionization, possibly from absorption of X-rays that originate in the primary ejecta from the down scattering of the γ-rays. Circumbinary gas emission is dwarfed by the continuum from the expanding photosphere but this outer reservoir of gas does produce numerous, relatively narrow absorption lines distributed throughout the continuum.

The impact of the primary ejecta on the circumbinary THEA gas at velocities >$10^3$ km/s creates forward and reverse shocks that heat both components of ejecta (see figures 1 and 3 of Weaver et al. 1977; also figure 4 of Contini & Prialnik 1997). The heated regions, comprising the circumbinary THEA gas for the forward shock and the primary WD ejecta for the reverse shock, having highest densities will dominate the post-shock

7emission spectrum. Depending upon the parameters of the circumbinary shell and the WD ejecta the emission-line spectrum for different novae will consist predominantly of gas that may (WD ejecta) or may not (circumbinary reservoir) have been processed by the nova outburst. This is important for the interpretation of the outburst based on emission line analyses since the outburst-processed layers of the WD will be mixed with gas from the secondary star that will have undergone its own evolutionary process. Figure 2 presents a schematic representation of the pre- and post-collision model proposed here for novae that follows from the Castor, McCray, & Weaver picture for hot stars.

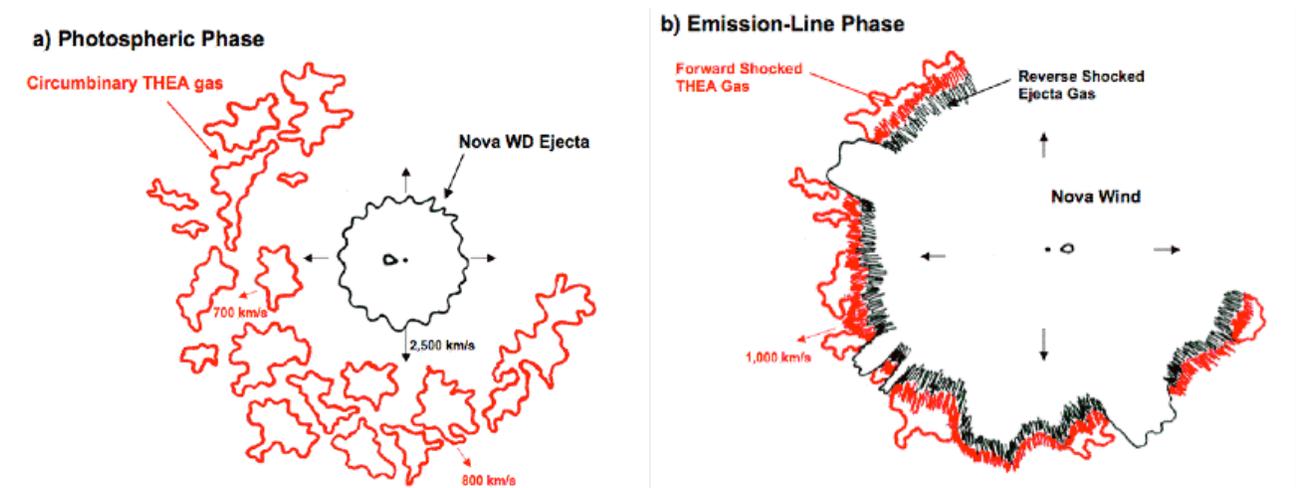

**Figure 2.** Schematic representation of (a) the photosphere phase before collision, and (b) the forbidden line phase with the shock heated gas after the collision.

### 4. Circumbinary gas properties from line profiles

The emission line profiles from high-resolution spectra of novae provide two important pieces of information about the circumbinary gas. Figure 3 shows a montage of emission profiles for a number of lines from novae in our survey from Paper I. We focus on the forbidden lines because they are optically thin, hence their profiles reflect the true velocity structure of the emitting gas as opposed to permitted lines that may be dominated by optical depth effects. A characteristic feature of the forbidden lines in most novae is that they tend to have rectangular, boxy flat topped profiles. Beals (1931) and Bappu & Menzel (1954) first drew attention to this characteristic and they showed that rectangular profiles result from optically thin emission in a uniformly expanding spherical shell. One cannot exclude the possibility that other combinations of geometries and density and velocity distributions may also produce rectangular profiles, nevertheless the rectangular profiles may be taken as evidence for an expanding shell of emitting gas. The impact of the WD ejecta on the circumbinary reservoir of gas should create just such a thin shell of shocked radiating gas that has spherical symmetry because THEA absorption systems are observed to exist along all lines of sight, i.e., have an approximate spherical distribution.

Velocity information from the absorption and emission lines associated with the colliding shells provides information on the relative masses of the primary and the circumbinary shells. The collision between the two systems releases kinetic energy that converts the spectrum from absorption to emission lines, whose profiles reflect the kinematical properties of the heated gas. The early evolution of the impacting shells is dominated by a momentum-conserving 'snowplow' phase (Steigman, Strittmatter, & Williams 1975; Castor, McCray, & Weaver 1975) where the primary ejecta sweep up circumbinary gas, creating a narrow interface of shocked gas. For most situations the gas in the forward shock, which is mostly THEA gas, will have a higher density than the reverse shocked gas of the primary WD ejecta due to the large decrease in density experienced by the WD ejecta from their expansion. Thus, the emission lines should originate in the forward shocked gas consisting predominantly of THEA gas mixed with some of the primary WD ejecta.

The widths of the forbidden lines are determined largely by the velocity of the gas in the forward shock. This velocity will vary throughout the duration of the collision as the momentum of the primary ejecta is transferred to the circumbinary gas, taking on values between the circumbinary and primary ejecta expansion velocities.

Conservation of momentum applied to the collision of the two shells requires that the column densities, $N_{cir}$ & $N_{ej}$, and velocities, $v_{cir}$ & $v_{ej}$, of the circumbinary and primary ejecta shells satisfy the condition

$$N_{cir}/N_{ej} = (v_{ej} - v_f)/(v_f - v_{cir}) , \qquad (3)$$

where $v_f$ is the final outward velocity of the merged systems. Pre-collision velocities can be determined from the absorption components of lines that originate in the two shells prior to the collision, before the forbidden emission lines appear in the spectrum. The final velocity is that corresponding to the widths of the forbidden lines weeks after they have appeared, after most of the WD ejecta from the outburst have impacted the circumbinary gas.

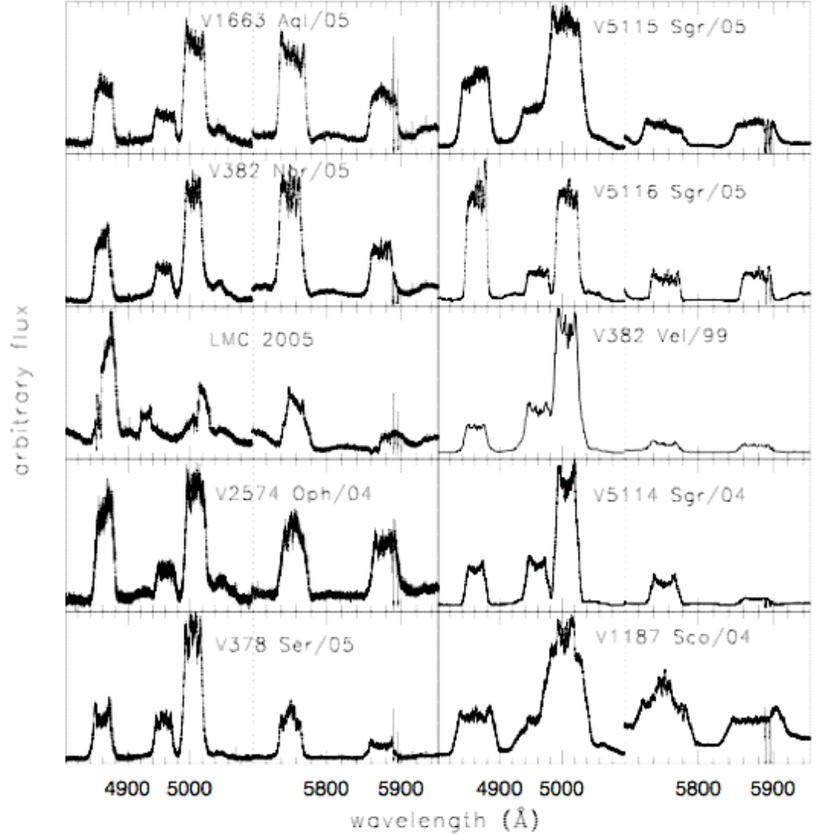

**Figure 3.** Profiles of the prominent Hβ, [O III] λλ4959, 5007, [N II] λ5755, & He I λ5876 lines during the emission-line phase of novae, months after outburst. The rectangular, or box-shaped, profiles of the forbidden lines are notable.

We have determined the relative velocities of the pre- and post-collision gas for the novae in our high resolution spectroscopic study. In table 2 we list the velocities of the primary WD ejecta and the circumbinary THEA gas taken from the absorption lines formed in each of the two components. The velocities change with time due to continuing acceleration of the gas so we have listed the measurements from the spectra at the last epoch before forbidden lines appear. The primary ejecta expansion velocity $v_{ej}$ has been determined from the broad Na I D absorption trough displacement taken relative to the central velocity of forbidden line emission determined at later epochs. The initial circumbinary velocity $v_{cir}$ has been determined from the Na I D wavelength shift of the THEA system(s). Some novae show multiple Na I D absorption doublets with different velocities, but there is usually only one system having sufficient column density to have detectable lines from Fe-peak and s-process elements. In two novae, V2574 Oph/04 and V2575 Oph/06, there were brief epochs when two separate THEA systems were observed, each with multiple absorption lines. The THEA velocities we list for these novae correspond to systems having comparable equivalent widths.

The post-collision velocity of the merged shells $v_f$ is determined from the widths of the forbidden lines that are excited by the collision of the two shells. We have measured the line widths of forbidden lines having a range of ionization when observed, e.g., [O III] $\lambda5007$, [N II] $\lambda5755$, and [O I] $\lambda6300$, and have determined the corresponding expansion velocity of the merged shells. In table 2 we list the velocity $v_f$ measured for the post-collision ejecta for each nova we have studied, corresponding to the mean of values obtained from the half-width at half maximum (HWHM) intensity of the above lines. There is a dispersion in the velocities measured from different lines in some of the novae due to ionization stratification, generally of order 150 km/s, so there is some uncertainty in the velocity assigned to an individual nova. Nevertheless, it is clear from table 2 that the final expansion velocity of the merged ejecta shells is much closer to that of the circumbinary THEA system than it is to the primary WD ejecta.

**Table 2.** Novae ejecta expansion velocities

| Nova | Expansion Velocities (km/s) | | |
|---|---|---|---|
| | Primary WD ejecta $V_{ej}$ | THEA system(s) $V_{cir}$ | Post-impact gas $V_f$ |
| V382 Vel/99 | 4,260 | N/A | 1,130 |
| V2573 Oph/03 | 2,640 | 710 | N/A |
| V5114 Sgr/04 | 2,570 | 1,020 | 1,010 |
| V2574 Oph/04 | 2,970 | 850; 990 | 880 |
| V1186 Sco/04 #1 | 2,540 | 440 | 480 |
| V1187 Sco/04 #2 | 2,010 | N/A | 1,920 |
| V382 Nor/05 | 3,620 | 810 | 910 |
| V378 Ser/05 | 2,230 | 680 | 860 |
| V5115 Sgr/05 #1 | 3,220 | N/A | 1,320 |
| V1663 Aql/05 | 2,190 | 890 | 910 |
| V5116 Sgr/05 #2 | 2,480 | 890 | 1,060 |
| V476 Sct/05 #1 | 2,370 | 570 | 670 |
| LMC 2005 | 1,630 | 450 | 610 |
| V2575 Oph/06 | 2,640 | 470; 840 | 580 |
| V5117 Sgr/06 | 2,630 | 820 | 710: |

For every nova for which circumbinary THEA gas has been observed the table 2 results show that the rapidly expanding primary WD ejecta are brought to a virtual halt when they collide with the circumbinary shell. This result also applies to two of the three novae for which no THEA systems were observed. THEA systems in those novae may have existed, but disappeared before our first observations because they were very short lived due to the high velocities of the primary ejecta. Only in V1187 Sco/04 do the emission-line nebular expansion velocities remain high, perhaps because of the absence of significant amounts of circumbinary gas. In all other objects, the result is inescapable: the column density of circumbinary gas, and therefore its mass, must substantially exceed that of the primary WD ejecta at the time the shells collide and create the nebular spectrum. This fact explains why novae ejecta at the time of outburst have velocities of thousands of km/s, whereas the expansion velocities of old, spatially resolved nova shells, years later, are always much smaller, at only hundreds of km/s. The rapidly expanding WD ejecta are strongly decelerated within the first months of the outburst from their impact with the more massive surrounding circumbinary gas.

**5. Summary**

High resolution spectra of novae enable a physical and kinematical picture to be constructed for novae ejecta. The key observational facts that point to a geometry of two colliding shells for the ejecta are (1) the superposition on the P Cygni profiles of narrow THEA absorption having a much lower velocity, (2) the acceleration of both the broad P Cygni and the narrow THEA circumbinary absorption features, which requires proximity of both components to the nova, (3) the relatively brief interval over which the early low ionization P Cygni absorption spectrum changes into the more highly ionized emission-line spectrum, (4) the much narrower widths of the emission-line spectrum compared to the preceding very broad P Cygni profiles, with the emission lines almost always having the same widths as the outer circumbinary THEA absorption displacements, and (5) the ubiquitous rectangular emission line profiles that are characteristic of emission from a relatively thin expanding spherical shell.

The spectra of novae in the weeks following maximum light give clear evidence for the existence of two distinct components of expanding gas. The broad P Cygni profiles originate in gas from the outer layers of the WD, ejected by the thermonuclear runaway at speeds >2,000 km/s. Narrow absorption lines from excited levels of the heavy elements having significantly lower expansion velocities are superposed upon the P Cygni emission components and continuum of the expanding photosphere. This circumbinary THEA gas is warm, at $T\sim10^4$ K with a low level of ionization, and must reside outside the expanding WD ejecta. The source of excitation is most likely UV radiation from the very hot WD during its run-up in temperature in the months/years prior to the TNR.

From the column densities derived in §2.1 and in Paper I for THEA systems we have analyzed the energy required to eject the gas at the observed velocities is of the order of the energy generated in a typical nova outburst. We hypothesize that the ejection of both components of gas occurs as a result of absorption of the radiant energy, e.g., X-rays from Comptonization of the γ-rays produced by the TNR. We argued in Paper I that the circumbinary THEA gas originated from pre-outburst secondary star episodic ejection events. We now accept that this is unlikely on energetic grounds because ejection episodes would require events that are comparable in energy to a nova outburst TNR, and should therefore be observed.

An alternative origin for a large reservoir of circumbinary gas exists that has been advocated by Sytov et al. (2007, 2009a) based on 3D hydrodynamical calculations of mass transfer in cataclysmic variables (CVs). Their numerical models produce an elliptical accretion disk that forms a bow shock from its interaction with a spiral tail of material that is lost through the outer L3 point. A spiral density wave of ejected material forms a common envelope with dense gas clumps having sizes the order of the binary separation (Sytov et al. 2009b). The spiral structure of the ejected gas is converted into a more circular geometry within a few rotations due to dissipative processes and the collisional interaction of the ejected gas in adjacent spirals (Sytov et al. 2009a). The resulting common envelope of circumbinary gas could be the reservoir of material that produces the observed THEA systems. The calculations indicate that this gas should have low ionization and significant column densities, and therefore should be detectable at high spectral resolution from Na I D and Ca II H&K absorption in quiescent cataclysmic variables. We are undertaking a study of such gas in quiescent CV systems to determine whether its existence is transitory near the time of outbursts or a stable long term phenomenon.

The existence of two distinct shells of gas has implications for dust formation in novae. The conventional picture of dust formation (Shore & Gehrz 2004) has it forming in the rapidly expanding ejecta from the WD. The shock produced by the collision of that ejecta with the outer circumbinary gas is likely to destroy any dust that forms before the collision of the two shells which converts the spectrum from the initial P Cygni absorption spectrum to the emission-line spectrum. However, subsequent radiative cooling of the swept up gas may provide another opportunity for dust to form, in which case it would be forming from a mixture of gas of both shells. In fact, dust may also form in the outer THEA gas before the collision of the two shells. It is conceivable that different types of dust may be present at the same time in each of the two expanding shells of gas prior to their merger. The formation of dust in novae may therefore depend on a number of parameters that include the abundances of the two shells and the timescale of the shell collision relative to the adiabatic and radiative cooling times of the ejecta.

The fact that THEA absorption is observed in virtually all novae indicates that the circumbinary envelope is not confined to just the plane of the binary systems. The mass ejection that occurs through the L3 point is largely confined to the equatorial plane of the binary system. Because of the appreciable internal energy of the gas, which derives from the collisional interaction of adjacent spiral streams driven by the orbital motion of the binary, the 3D hydrodynamical calculations show that there is some diffusion of ejected gas out of the binary orbital plane toward the poles of the orbit (Bisikalo 2009). Given the typical Fe II column densities of $\sim 10^{18}$ cm$^{-2}$ observed for THEA absorption, if the gas and column densities in the direction of the poles are $10^3$ times smaller than those in the orbital plane there would still be sufficient heavy element column densities in these directions to produce observable THEA absorption. However, the polar column densities may be lower than this value, in which case the observation of THEA absorption in the large majority of novae may be difficult to explain by the mechanism described here. Until more detailed calculations of the L3 mass ejection are performed that demonstrate that circumbinary gas does migrate out of the binary orbital plane, the paradigm proposed here may be problematical in explaining the existence of THEA systems in the majority of novae.

    The clear presence of THEA systems at the time of nova outbursts raises the question of the possible role of the circumbinary gas in outbursts. Dwarf novae outbursts are understood to be caused by a release of gravitational energy from partial collapse of the accretion disk in CVs (Warner 1995). The fact that almost all classical novae in our survey are observed to be associated with a large reservoir of circumbinary gas suggests that there may be a similar causal relationship. The existence of a large common envelope is very suggestive that some classical novae outbursts may be triggered not by steady accretion of mass via the inner L1 Lagrange point but by the collapse of a portion of the circumbinary reservoir onto the WD, as occurs for the dwarf novae outbursts. This process might explain the secondary maxima that are observed in many novae, which are difficult to understand in the conventional picture of steady L1 mass transfer.

    The interpretation of the spectral evolution of novae in terms of two discrete colliding shells of gas (1) explains the relatively rapid transition of the early absorption spectrum to a forbidden emission-line spectrum, (2) may dictate how dust forms in the ejecta, depending on the timing of the collision relative to other factors (Gallagher 1977; Shore & Gehrz 2004), (3) may help explain the complex behavior of X-ray emission, which can strengthen weeks or months after the outburst (Ness et al. 2007), (4) provides for the creation of a thin shell of radiating gas from the resulting shock that has spherical symmetry, consistent with the rectangular emission-line profiles observed in novae, and (5) provides a natural explanation for the relatively narrow emission-line widths compared to the high pre-collision absorption velocities of the primary ejecta.

    Specifying the geometry of the emitting gas in novae is crucial to interpreting their spectra. If further analysis substantiates the colliding shells of ejecta it will provide a more solid basis for the interpretation of spectra for all wavelength regions. One of the important consequences of the picture presented here is that the merger of the primary and THEA systems excites the nebular forbidden line spectrum. The gas is therefore a mixture of ejecta having different histories, so abundances derived from the emission lines do not necessarily reflect those of the material that underwent the TNR.

## 6. References

Arnett W D 1979 *ApJ* **230** L37
Bappu M K V and Menzel D H 1954 *ApJ* **109** 508
Beals C S 1931 *MNRAS* **91** 966
Bisikalo D V 2009, presentation at Cataclysmic Variables Workshop, Tucson, AZ
Cassatella A, Lamers H J G L M Rossi C, Altamore A, and González-Riestra R 2004 *A&A* **420** 571
Castor J, McCray R and Weaver R 1975 *ApJ* **200** L107
Contini M and Prialnik D 1997 *ApJ* **475** 803
Fruchter A, Krolik J H and Rhoads J E 2001 *ApJ* **563** 597
Gallagher J S 1977 *AJ* **82** 209
Gehrz R D, Truran J W, Williams R E and Starrfield S 1998 *PASP* **110** 3
Glasner S A, Livne E and Truran J W 2007 *ApJ* **665** 1321
Gomez-Gomar J, Hernanz M, Jose J and Isern J 1998 *MNRAS* **296** 913
Hauschildt P H, Shore S N, Schwarz G J, Baron E, Starrfield S and Allard F 1997 *ApJ* **490** 803
Jenkins E B, Bowen D V, Tripp T M and Sembach K R 2005 *ApJ* **623** 767
Kato M and Hachisu I 2007 *ApJ* **657** 1004
------------ 2009 *ApJ* **699** 1293
Livio M, Mastichiadis A, Ogelman H and Truran, J W 1992 *ApJ* **394** 217
McLaughlin D B 1943 *Pub. Mich. Obs.* **8** 149
------------- 1947 *PASP* **59** 81
Mihalas D 1978 *Stellar Atmospheres* 2$^{nd}$ ed (San Francisco: W H Freeman)

Ness J-U, Schwarz G J, Retter A, Starrfield S, Schmitt J H M M, Gehrels N, Burrows D and Osborne J 2007 *ApJ* **663** 505
Orio M 1999 *Phys Repts* **311** 419
Rosseland S 1946 *ApJ* **104** 324
Russell H N 1936 *PASP* **48** 29
Schwarz G J, Shore S N, Starrfield S, Hauschildt P H, Della Valle M and Baron E 2001 *MNRAS* **320** 103
Senziani F, Skinner G K, Jean P and Hernanz M 2008 *A&A* **485** 223
Shore S N and Gehrz R D 2004 *A&A* **417** 695
Shore S N et al 2003 *AJ* **125** 1507
Short C I, Hauschildt P H, Starrfield S and Baron E 2001 *ApJ* **547** 1057
Starrfield S G, Truran J W and Sparks W M 2000 *New Astr Repts* **44** 81
Steigman G, Strittmatter P A and Williams R E 1975 *ApJ* **198** 575
Sytov A Y, Bisikalo D V, Kaigorodov P V and Boyarchuk A A 2009a *Astr Repts* **53** 223
\_\_\_\_\_\_ 2009b *Astr Repts* **53** 428
Sytov A Y, Kaigorodov P V, Bisikalo D V, Kuznetsov O A and Boyarchuk A A 2007 *Astr Repts* **51** 836
Warner B 1995 *Cataclysmic Variable Stars* (New York: Cambridge Univ. Press)
Weaver R, McCray R, Castor J, Shapiro P and Moore R 1977 *ApJ* **218** 377
Williams R E 1972 ApJ **178** 105
Williams R E, Mason E, Della Valle M and Ederoclite A 2008 *ApJ* **685** 451 (Paper I)
Yaron O, Prialnik D, Shara M M and Kovetz A 2005 *ApJ* **623** 398